\documentclass[pre,aps,showpacs,twocolumn]{revtex4}
\usepackage{amsfonts}
\usepackage{amsmath}
\usepackage{amssymb}
\usepackage{graphicx}
\usepackage{epsfig}
\usepackage{bm}
\begin{document}

\title{Imperfect Linear Optical Photonic Gates with Number-Resolving Photodetection}

\author{A. Matthew Smith$^{1,2}$, D. B. Uskov$^{1,3}$, L. H. Ying$^1$, and L. Kaplan$^1$}
\affiliation{$^1$Department of Physics and Engineering Physics, Tulane University, New Orleans, Louisiana 70118, USA  \\
$^2$Air Force Research Lab, Information Directorate, Rome, New York 13440, USA \\
$^3$Department of Mathematics and Natural Sciences, Brescia University, Owensboro, Kentucky 42301, USA}

\begin{abstract}
We use the numerical optimization techniques of Uskov {\it et al.}~\cite{uskov1} to investigate the behavior of the success rates for KLM style~\cite{klm} two- and three-qubit entangling gates.  The methods are first demonstrated at perfect fidelity, and then extended to imperfect gates.  We find that as the perfect fidelity condition is relaxed, the maximum attainable success rates increase in a predictable fashion depending on the size of the system, and we compare that rate of increase for several gates.
\end{abstract}


\maketitle

\section{Introduction}
Optics provides one of the most promising implementations for quantum information processing, due to long photon decoherence times, the ease of photon manipulation, and the ability to transmit quantum state information over very large distances. Optics have also been suggested as buses in hybrid matter-optical quantum computers.  It is therefore desirable to be able to efficiently create and manipulate states at the single photon level. In this work, we will focus on manipulating states through optical gate implementation, rather than on single photon or quantum state generation~\cite{van}.

Any number of implementation schemes have been proposed to preform non-trivial operations at the single photon level.  These include non-linear materials, coupling through atomic interactions, Zeno-type non-unitary interactions~\cite{franson}, and silica-on-silicon waveguides~\cite{o'brien}.

Knill, Laflamme, and Milburn significantly advanced the prospect of single photon quantum computing in their seminal paper~\cite{klm}, in which they overcame the need for nonlinear interactions by using the inherent nonlinearity of photon measurements.  In this scheme, the computational system is combined with ancillary modes, and the gate operation is performed on the enlarged state space. red states of the computational system when the measurement is successful).  The ancilla modes are measured with photon-number-resolving detectors, leaving the computational modes undisturbed and in the desired output state provided the measurement is successful. The probabilistic nature of quantum measurement implies a trade-off between the success rate of the operation (the probability of obtaining the desired measurement outcome for the ancillary modes) and the fidelity (the overlap between the actual and desi

In previous work~\cite{uskov1}, the authors have shown that a combination of analytical and numerical techniques may be used to design optimal linear optics transformations implementing two- and three-qubit entangling gates. Specifically, solutions were obtained that have the maximum possible measurement success probability given the constraint of perfect fidelity for a given desired gate. In practical implementations, however, the goal of perfect fidelity may not always be desirable or even attainable (for given ancilla resources). In the present work, we therefore generalize our previous techniques to the case of imperfect fidelity, and investigate the above-mentioned trade-off between the fidelity and success of the linear optical transformation. We will show that for sufficiently small deviation from perfect fidelity, a single parameter determines the relationship between fidelity and optimal success rate. We will also observe that once the perfect fidelity condition is relaxed, the resulting relative gain in success rate appears to grow with the size of the linear optical system under consideration.

\section{LOQC Gate Optimization}
\begin{figure}[htbp]
   \centerline{\includegraphics[width=0.42\textwidth]{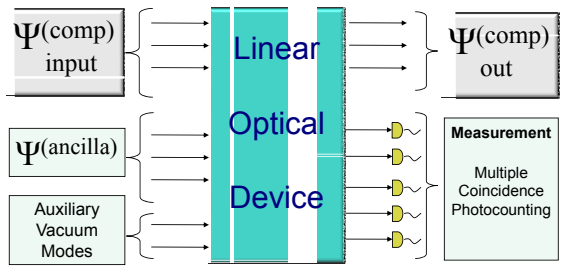} }
   \caption{(Color Online) A scheme for an LOQC transformation or a quantum state generator~\cite{uskov1}. The computational input state is a separable state of two or more dual-rail encoded qubits. The ancilla state may be separable or entangled.}
\label{figschematic}
\end{figure}

A Linear Optical Quantum Computing (LOQC) measurement-assisted transformation, as suggested by KLM, is schematically illustrated in Fig.~\ref{figschematic}.
Here the input state $|{\Psi}_{\rm in}^{\rm comp}\rangle\times| \Psi^{\rm ancilla}\rangle$ is a product of the computational state and an ancilla state, with $|{\Psi}_{\rm in}^{\rm comp}\rangle$ containing $M_c$ photons in $N_c$ modes, and $| \Psi^{\rm ancilla}\rangle$, containing $M_a$ photons in $N_a$ modes. 

The $N_c$ computational modes are those on which the actual gate is intended to act.  Assuming the Fock basis and dual-rail encoding, each qubit is represented by a single photon in two computational modes: specifically the logical states $|\!\uparrow\rangle$ and $|\!\downarrow\rangle$
are represented by $|1_1,0_2 \rangle$ and $|0_1,1_2 \rangle$, respectively.   The number of computational qubits is then equal to the number of photons $M_c$ in the input state, which is half the number of modes: $M_c=N_c/2$.  This also implies that any output state with more or fewer than one photon in any logical qubit is not a computationally valid output.
 
In Sec.~\ref{sec2qubit}, we first discuss two-qubit gates, so the computational state will consist of $M_c=2$ photons in $N_c=4$ modes.
However, our numerical optimization approach applies to any gate operation~\cite{uskov1}, e.g., the three-qubit Toffoli gate investigated in Sec.~\ref{sectoffoli}, where the computational state consists of $M_c=3$ photons in $N_c=6$ modes, and others.

The ancilla state may be chosen from a wide array of possible states, and this state may be separable, entangled, or an ebit state carrying spatially distributed entanglement~\cite{wilde}. We note that there is a lower limit to the amount of ancilla resources needed to implement any given gate. Additionally, we allow for the possibility of an arbitrary number $N_v$ of vacuum modes that are unoccupied in the input state.

The linear optical device illustrated in Fig.~\ref{figschematic} transforms a set of input creation operators $a_i^{(\rm in)\dag}$ to a weighted sum of output creation operators $\sum_j U_{i,j} a_j^{(\rm out)\dag}$. Here $U$, which contains all physical properties of the device, is an $N \times N$ unitary matrix, where $N=N_c+N_a+N_v$ is the total number of modes (computational, ancilla, and vacuum).
The total input state may be written in the Fock representation as $|\Psi_{\rm in} \rangle = |n_{1}, n_2, \ldots, n_N \rangle$, where $n_i$ is the occupation number of the $i$-th input mode, and $\sum n_i=M_c+M_a=M$ is the total number of photons.  The transformation $\hat {\Omega}$ then takes the form
\begin{equation}
\label{Eq:OutState}
|\Psi_{\rm out}\rangle = \hat{\Omega} |\Psi_{\rm in}\rangle = \prod_{i=1}^{N} \frac{1}{\sqrt{n_{i} !}} \left( \sum_{j=1}^N U_{i,j} a_{j}^{(\rm out)\dag} \right)^{n_i}  |0 \rangle \,.
\end{equation}
We note that
$\hat {\Omega}$ is a multivariate polynomial of degree $M$ in the elements $U_{i,j}$.

Once the transformation is complete, a measurement is applied to the $N_a+N_v$ ancilla and vacuum modes.
This measurement can be formally described by a Kraus POVM~\cite{kraus} operator acting on these modes only: $\hat{P} = |0_{N_c+1}, 0_{N_c+2},\ldots, 0_{N}\rangle \langle \Psi_{\rm measured}|$.
In the case of a number-resolving photon-counting measurement, $\langle \Psi_{\rm measured}|=\langle k_{N_c+1}, k_{N_c+2}, \ldots, k_{N}| $, where $k_i$ is the number of photons measured in the $i$-th mode of the ancilla.  Finally, the resulting transformation of the computational state is a contraction quantum map  $|\Psi_{\rm out}^{\rm comp} \rangle = \hat{A} |\Psi_{\rm in}^{\rm comp} \rangle/\| \hat{A} |\Psi_{\rm in}^{\rm comp} \rangle\|$~\cite{kraus}, where $\hat{A}=\hat{A}(U)$ is defined by
\begin{equation}
\label{Eq:proj}
\hat{A} | \Psi_{\rm in}^{\rm comp} \rangle = \langle k_{N_c+1}, k_{N_c+2}, \ldots, k_{N}| {U} |\Psi_{\rm in}\rangle\,.
\end{equation}
The linear operator $\hat{A}$ contains all the information of relevance to the gate or state transformation.
Again the entries of the $\hat{A}$ matrix, $\hat A_{i,j}=\hat A_{i,j}(U)$ are multivariate polynomials of degree $M$ in the entries of $U$.

 We note that the dual-rail computational basis is a subset of all possible states of $M_c$ photons in the $2M_c$ computational modes. Since the input state of the computational modes necessarily belongs to the computational space, while the output state may in general be any state of $M_c$ photons in $2M_c$ modes, the transformation matrix $\hat{A}$ is a rectangular matrix, mapping the computational Hilbert space, of dimension $2^{M_c}$, to a larger Hilbert space, of dimension $(3M_c-1)!/(2M_c-1)!(M_c)!$.  For example, $\hat{A}$ is a $10 \times 4$ matrix for a two-qubit gate.
Similarly $\hat{A}$ is a $56 \times 8$ matrix for a three-qubit gate and $35 \times 9$ for two-mode biphotonic qutrit gates~\cite{lanyon}.

We now define precisely the operational fidelity of a transformation, which in general differs from the common measure of fidelity for a state transformation~\cite{nielsen}. Physically, the transformation $\hat{A}$ has $100 \% $  fidelity  if it is proportional to the target transformation $\hat{A}^{\rm Tar}$, i.e., $\hat{A} = g \hat{A}^{\rm Tar}$, where $g$ is an arbitrary complex number (in which case $S= |g|^2$ is the success probability of the transformation~\cite{lapaire}). In general, we may consider complex rays $\beta_1\hat{A} $ and $\beta_2\hat{A}^{\rm Tar} $,  $\beta_1 , \beta_2 \subset \mathbb{C} $, as elements of a complex projective space, and define the fidelity as
\begin{equation}
\label{Eq:F}
F(\hat A)=\frac{\langle \hat{A} | \hat{A}^{\rm Tar} \rangle \langle \hat{A}^{\rm Tar} | \hat{A} \rangle}{\langle \hat{A} | \hat{A} \rangle \langle \hat{A}^{\rm Tar} | \hat{A}^{\rm Tar}\rangle } \,,
\end{equation}
where the Hermitian inner product is $\langle \hat A | \hat B \rangle \equiv {\rm Tr}(\hat A^{\dag} \hat B )/D_c$, and $D_c=2^{M_c}$ is the dimension of the computational space. $F$ is closely related to the Fubini-Study distance $\gamma=\cos^{-1}(\sqrt{F})$~\cite{bengtsson}, but for numerical computations $F$ has the advantage of being non-singular near $F=1$.

In general, the success probability $S$ depends on the initial state $|\Psi_{\rm in}^{\rm comp} \rangle$. $S$ is bounded above by the square of the operator norm, $\| \hat A \|^2 \equiv (\| \hat{A}\|^{\rm Max})^2 = {\rm Max}(\langle \Psi_{\rm in}^{\rm comp} | {\hat A}^{\dag} \hat A| \Psi_{\rm in}^{\rm comp} \rangle)$, and below by $(\| \hat A \|^{\rm min})^2 = {\rm Min}  (\langle \Psi_{\rm in}^{\rm comp} | {\hat A}^{\dag} \hat A | \Psi_{\rm in}^{\rm comp} \rangle)$, where the maximum and minimum are taken over the set of properly normalized input states. As a more convenient measure, we use the Hilbert-Schmidt norm $\langle\hat{A} |\hat{A}\rangle $. It is easy to verify that $(\| \hat{A}\|^{\rm Min})^2 \leq 
\langle\hat{A} |\hat{A}\rangle \leq (\| \hat{A}\|^{\rm Max})^2 $. As fidelity $F \rightarrow 1$, $ \| \hat{A}\|^{\rm Min}/\| \hat{A}\|^{\rm Max} \rightarrow 1$ and $S$ becomes a well-defined quantity equal to the Hilbert-Schmidt norm. In the following, we refer to
\begin{equation}
\label{Eq:S}
S(\hat A)=\langle\hat{A} |\hat{A}\rangle
\end{equation}
 as {\it the} success probability, keeping in mind that it may not correspond to the success probability for every initial state, except in the case of perfect fidelity. Since the state transformation $\hat A$ is a function of the linear mode transformation $U$, the fidelity $F$ and success $S$ are also functions of $U$.

In the above discussion, we have assumed unitarity of the underlying linear mode transformation $U$. However, an arbitrary $(N_c+N_a) \times (N_c+N_a)$ complex matrix $U$ of unit spectral norm ($\| U \|=  1$, where $\| U\|$ is the largest singular value of $U$) may be shown via the unitary dilation technique~\cite{knill,uskov2} to be equivalent to an $(N_c+N_a+N_v) \times  (N_c+N_a+N_v)$ unitary matrix having the same success and fidelity, where $N_v$ is the number of vacuum modes in the input and output states. Thus, it is very convenient in practice to relax the unitarity condition and consider general matrices $U$ of norm $1$, with the understanding that the number of singular values in $U$ different from unity corresponds to the number of vacuum modes that are required to implement that solution~\cite{uskov2}. Furthermore, we may consider arbitrary complex matrices $U$ with the rescaling $U \to U/\| U\|$ to ensure unit norm. The fidelity function (\ref{Eq:F}) is unaffected by the rescaling, while the success function (\ref{Eq:S}) must be generalized as
\begin{equation}
\label{Eq:Sgen}
S(\hat A)=\langle\hat{A} |\hat{A}\rangle/ \|U\|^{2M}
\end{equation}
for general complex $U$. The success $S(\hat A)$ exhibits cusp-like singular behavior in this extended search space whenever the largest singular value of U goes through a double or higher-order degeneracy. This occurs in particular in the neighborhood of the manifold $U^\dagger U=I$.

The optimization problem we address is to maximize the success probability $S$ for a given target transformation ${\hat A}^{\rm Tar}$, for given ancilla resources, and for a given fidelity level $F \le 1$. This is mathematically equivalent to unconstrained maximization of the function $S+F/\epsilon$ in the space of all $N \times N$ matrices $U$, where $1/\epsilon$ is a Lagrange multiplier. Here $\epsilon \to 0^+$ corresponds to maximizing the success probability while requiring perfect fidelity ($F=1$). As $\epsilon$ is increased, the maximum of $S+F/\epsilon$ yields linear optics transformations that maximize the success $S$ as a function of the fidelity $F$.

Given one optimal transformation $U$ that (locally or globally) maximizes success $S$ for a given fidelity $F$, $\epsilon$ may in general be continuously varied to obtain a one-parameter family of optimal transformations, tracing out a curve in success-fidelity space. Multiple such families of solutions may coexist, corresponding to different local maxima of the success rate at a single given value of the fidelity.

%discussed with the {\sc{cnot}} gate
%Furthermore, the optimal curves may intersect, so that the family of solutions obtained by deforming the optimal $F=1$ solution may cease to be globally optimal once the fidelity deviates sufficiently from unity.

Numerically, as we attempt to follow a family of optimal solutions, we find that it is necessary to perturb the $U$ matrix slightly at each step before incrementing $\epsilon$ and repeating the optimization process. This is due to the fact that such a family is typically  associated with a degeneracy of several, or in some cases all, singular values of $U$. The family of optimal solutions then
exists along a multi-dimensional ridge of the success function $S$ (Eq.~(\ref{Eq:Sgen})), with cusplike behavior of $S$ and thus of the optimization function $S+F/\epsilon$ along directions in $U$ space orthogonal to the ridge orientation. A slight perturbation of the $U$ matrix at each successive value of $\epsilon$ moves the solution away from the cusp and allows the ridge to be followed in a continuous manner.

%Theoretically this perturbation is not needed however due to finite precision we find that the gradient along this curve is at times too small to detect, there can also be local maximums of the optimization function present. By perturbing the system slightly, we move away from the desired curve to points with larger gradients allowing us to optimize in the direction $S+(F/\epsilon)$, sometime referred to as tunneling through higher dimensions. The matrix corresponding to the optimal point for any optimization step was then used as the initial condition for the next step, with a slightly different optimization function, ie  incremented $\epsilon$.  %We found empirically that a perturbation approach also helps to avoid the many local maximums in the in the system.  We parametrize each of the element of the complex matrix $U$, with elements of the form $r_ie^{\phi_i}$.  We chose a simple random in-phase kick to several elements of the matrix $U$, as $U$ is not strictly unitary anyway.  Without loss of generality we  can randomize several $r_i$.  

\section{Two-Qubit Gates}
\label{sec2qubit}
In this section we examine two controlled-phase gates, {\sc{cnot}} ($\theta=180$) and  {\sc{cs}}($90^\circ$) ($\theta=90$), which are of course equivalent to more general controlled-unitary gates by local qubit transformations. We also consider the more general $B$ gate, which is notable because an arbitrary two-qubit $SU(4)$ gate may be constructed out of two $B$ gates and local rotations, as shown by Zhang et al.~\cite{zhang}.  The maximum success rates for the two phase gates was first determined numerically by E. Knill~\cite{knill}.  Subsequently the maximum success rates of all of the controlled-phase gates and most general gates were found numerically by the authors \cite{uskov1}.  These solutions were found at perfect fidelity.  A question that remains to be answered is: what are the maximum success rates for imperfect gates with fidelity (slightly) less than 1?

We use the Cartan decomposition KAK to define each of the target transformations ({\sc{cs}}($90^\circ$), {\sc{cnot}}, and the B gate) using only three real parameters $(c_1,c_2,c_3)$.  For details on the Cartan decomposition see~\cite{zhang,khaneja} and \cite{uskov1}. In this decomposition, {\sc{cnot}} (also equivalent to {\sc{cz}} via local rotations) is represented by the point $(\pi/2,0,0)$, {\sc{cs}}($90^\circ$) is represented by $(\pi/4,0,0)$, and the B gate is represented by$(\pi/2,\pi/4,0)$.
 
 \begin{figure}[htbp]
   \centerline{\includegraphics[width=0.42\textwidth]{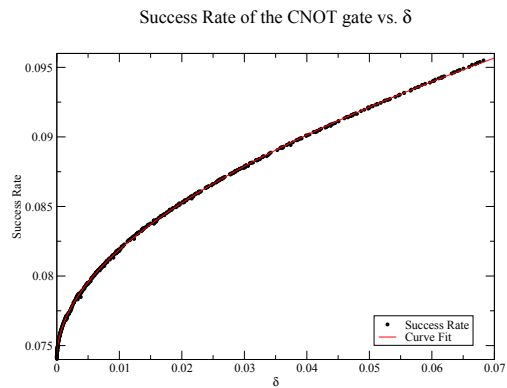} }
   \caption{(Color Online) Numerical results for the success rate $S$ of the {\sc{cnot}} gate as a function of $\delta=1-F$ for our perturbative relaxation of the $F=1$ constraint. The continuous curve is a fit given by Eq.~(\ref{s-numbers-cnot}).}
\label{cnot-fig}
\end{figure}

We begin with the {\sc{cnot}} gate as shown in Fig.~\ref{cnot-fig}. We note that $M_c=2$ computational photons in $N_c=4$ computational modes are required for any two-qubit gate in the dual rail representation. Additionally, the minimum ancillary resource required to implement {\sc{cnot}} with perfect fidelity consists of two single-photon ancillas, i.e. the {\sc{cnot}} gate requires $M_a=2$ ancilla photons in $N_a=2$ ancilla modes~\cite{uskov2}. Therefore we have a total of four photons in six modes in the initial state, and the linear optical transformation $U$ is a $6 \times 6$ matrix. Since the fidelity $F$ is a smooth function of $U$, the fidelity behaves quadratically, $F=1-O(\eta^2)$, around the maximum $F=1$, while the success $S$ is in general not at a local maximum at the $F=1$ point and therefore behaves generically as $S = S_0+a \eta+b \eta^2+\cdots$. Here $\eta$ is a local coordinate in $U$ space along the direction of the gradient of $S$ (with $\eta=0$ corresponding to a solution with perfect fidelity), $S_0$ is the success rate for perfect fidelity, and $a=|\nabla S|$. We then immediately see that the curve of success vs. fidelity should be of the form,
\begin{equation}
\label{s-funct1}
S(F)=S_0+S_1(1-F)^{1/2}+S_2(1-F)+ \cdots
\end{equation}
in the area around $F=1$, i.e., a power series expansion in  $(1-F)^{1/2}$. Defining $\delta=1-F$, we have
\begin{equation}
\label{s-funct}
S(\delta)=S_0+S_1\delta^{1/2}+S_2\delta +\cdots \,.
\end{equation}
Looking at Fig.~\ref{cnot-fig}, we see that the data from our numerical optimization does follow this form very well, with a best fit
\begin{equation}
\label{s-numbers-cnot}
S(\delta)=0.0740+0.0765\,\delta^{1/2}+0.0199\,\delta \,.
\end{equation}

When $F=1$, $\delta=0$ and the function $S(\delta=0)$ reduces to the $S=2/27\approx 0.0740$ result found by Knill~\cite{knill}, which was confirmed previously by the authors~\cite{uskov1}.  The ratio ${S_1/S_0}=1.03$ contains the most interesting information about the system, as it is a measure of the relative increase in success as the $F=1$ constraint is relaxed. We will use this ratio to compare the behavior of several gates.  We note that the matrices $U$ corresponding to the curve in Fig.~\ref{cnot-fig} remain unitary along the entire length of the curve and maintain the ``Knill Form.''  The ``Knill Form'' is a empirical statement about the structure of the $U$ matrix for optimal success with perfect fidelity in the case of the controlled-unitary and Toffoli gates, where one mode in each qubit may be chosen to be non-interacting (up to local rotations)~\cite{knill,uskov1,uskov2}.  

\begin{figure}[htbp]
   \centerline{\includegraphics[width=0.42\textwidth]{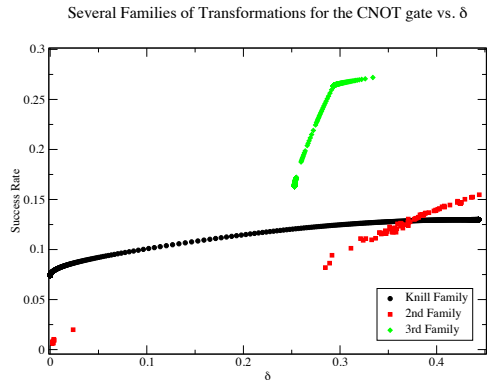} }
   \caption{(Color Online) The numerical results of fidelity-success optimization for the {\sc{cnot}} gate, extended to a larger domain of $\delta=1-F$ than in Fig.~\ref{cnot-fig}, and showing two additional one-parameter families of solutions that cross or surpass the ``Knill Family''  of solutions at finite $\delta$.}
\label{fam-fig}
\end{figure}

We also note that as mentioned above, this curve obtained by continuously deforming the $F=1$ optimal solution is not guaranteed to represent the {\it global} maximum of the success rate for all values of $F$. Indeed it is not surprising that for large deviations from perfect fidelity,  $\delta \equiv 1-F\geq .25$, we may obtain solutions with higher success rate $S$ for the same fidelity $F$, corresponding to other families of 
optimal solutions. Fig.~\ref{fam-fig} shows this for the case of the {\sc{cnot}} gate, in which we find two other families of solutions. One of these has success rates significantly higher than the ``Knill Form'' solution when $\delta\geq .25$ ($F \leq 0.75$), and the other has a fidelity-success curve that crosses that of the ``Knill Form'' family.  While these results are interesting, the large $\delta$ values corresponding to fidelities $F \le 0.75$ make these alternative families of solutions less useful in practice as gates.

As discussed in the previous section, the success rates shown in Fig.~\ref{fam-fig} and in all following figures are given by the Hilbert-Schmidt norm, i.e., these are the success rates averaged over all possible computational input states. We find that while the spread between the minimum and maximum success rates is $\le 10^{-4}$ for the 3rd family of solutions, the corresponding spread for the Knill family and 2nd family of solutions is of the same order as the average success rate. This suggests that in applications where single quantum state generation or transformation of a single input state is desired, the method described here may be used to design a linear optical device that maximizes the success rate for that specific application.

 \begin{figure}[htbp]
   \centerline{\includegraphics[width=0.42\textwidth]{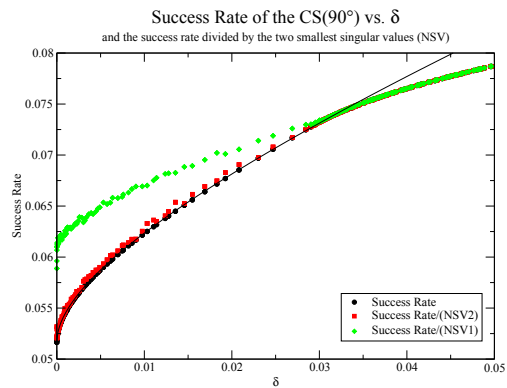} }
   \caption{(Color Online) Black circles show numerical results for the success rate $S$ as a function of $\delta=1-F$ for the {\sc{cs}}($90^\circ$) gate, obtained by continuously deforming the optimal $F=1$ solution. The solid curve (Eq.~(\ref{s-numb-cs90}))  is a best fit to the data. The green diamonds and red squares show the success rate divided by the normalized smallest and second smallest singular values, respectively.}
\label{cs90-fig}
\end{figure}

The same logic that describes the shape of the fidelity-success curve for {\sc{cnot}} holds true for any LOQC gate, including {\sc{cs}}($90^\circ$), the B gate, and the Toffoli gate.  In Fig.~\ref{cs90-fig} we show several results for {\sc{cs}}($90^\circ)$. The black circles show the success rates for {\sc{cs}}($90^\circ$) found using our perturbative optimization method. The solid curve is the best fit to the theoretical form (\ref{s-funct}),  and has the values
\begin{equation}
\label{s-numb-cs90}
S(\delta)=0.05165+.08368\,\delta^{1/2}+0.23282\,\delta \,.
\end{equation}
As in the case of {\sc{cnot}}, the $\delta=0$ point confirms results found previously \cite{knill,uskov1}.  The ratio $S_1/S_0=1.62$ for the {\sc{cs}}($90^\circ$) gate is larger than that of the {\sc{cnot}} gate.  

In contrast with the {\sc{cnot}} solution, where the optimal linear transformations $U$ associated with the Knill family of solutions are all unitary,
the optimal $(N_c+N_a) \times  (N_c+N_a) = 6 \times 6$ matrix for the {\sc{cs}}($90^\circ$) gate at $F=1$ ($\delta=0$) has a four-fold degenerate maximum singular value and two smaller non-degenerate singular values~\cite{knill}. As discussed above, the maximum singular value is always normalized to unity.
This means in practice that the physical device implementing this optimal solution requires two additional vacuum modes for unitary dilation~\cite{uskov1,uskov2,van}, for a total number of modes $N_c+N_a+N_v=8$.

The behavior of the non-degenerate singular values when the perfect fidelity condition is relaxed can be seen in Fig.~\ref{cs90-fig}. 
As noted earlier, the black circles show the maximal success rate at each value of the fidelity, which may also be thought of as the maximal success rate divided by the {\it largest} normalized singular value (NSV), which is always normalized to unity. The green diamonds show the success rate divided by the {\it smallest} NSV at each point, and similarly, the red squares show the success rate divided by the {\it second smallest} NSV.  Here we observe interesting behavior: as we move away from the perfect fidelity solutions by increasing $\delta$, the singular values converge. Now the second smallest NSV starts out within a few percent of unity, so it is difficult to determine exactly where it merges with the maximal NSV.  The data associated with the smallest NSV is more illustrative: the green diamonds show that the smallest NSV merges with the other singular values at $\delta\approx 0.03$.  At this point the optimal solution becomes unitary, and it remains unitary for larger $\delta$, as indicated by the three data sets overlapping.  This means that not only is our method optimizing the success rate at a given fidelity; it is actually optimizing the system size as well, as unitary dilation is no longer needed beyond the transition point. This reduction in system size and singular value convergence will be seen again in other gates.

We remark that while the perturbative expansion (\ref{s-funct}) works very well up until the NSV coalescence point at $\delta \approx 0.03$, it clearly fails at larger values of $\delta$. This is consistent with the fact that the structure of the optimal matrix $U$ is different on the two sides of the transition. The solutions with different degeneracy of the singular value spectrum should be regarded as constituting separate families of optimal transformations, although the optimal success rate behaves continuously at the coalescence point, where a transition between the two families occurs.  

\begin{figure}[htbp]
   \centerline{\includegraphics[width=0.42\textwidth]{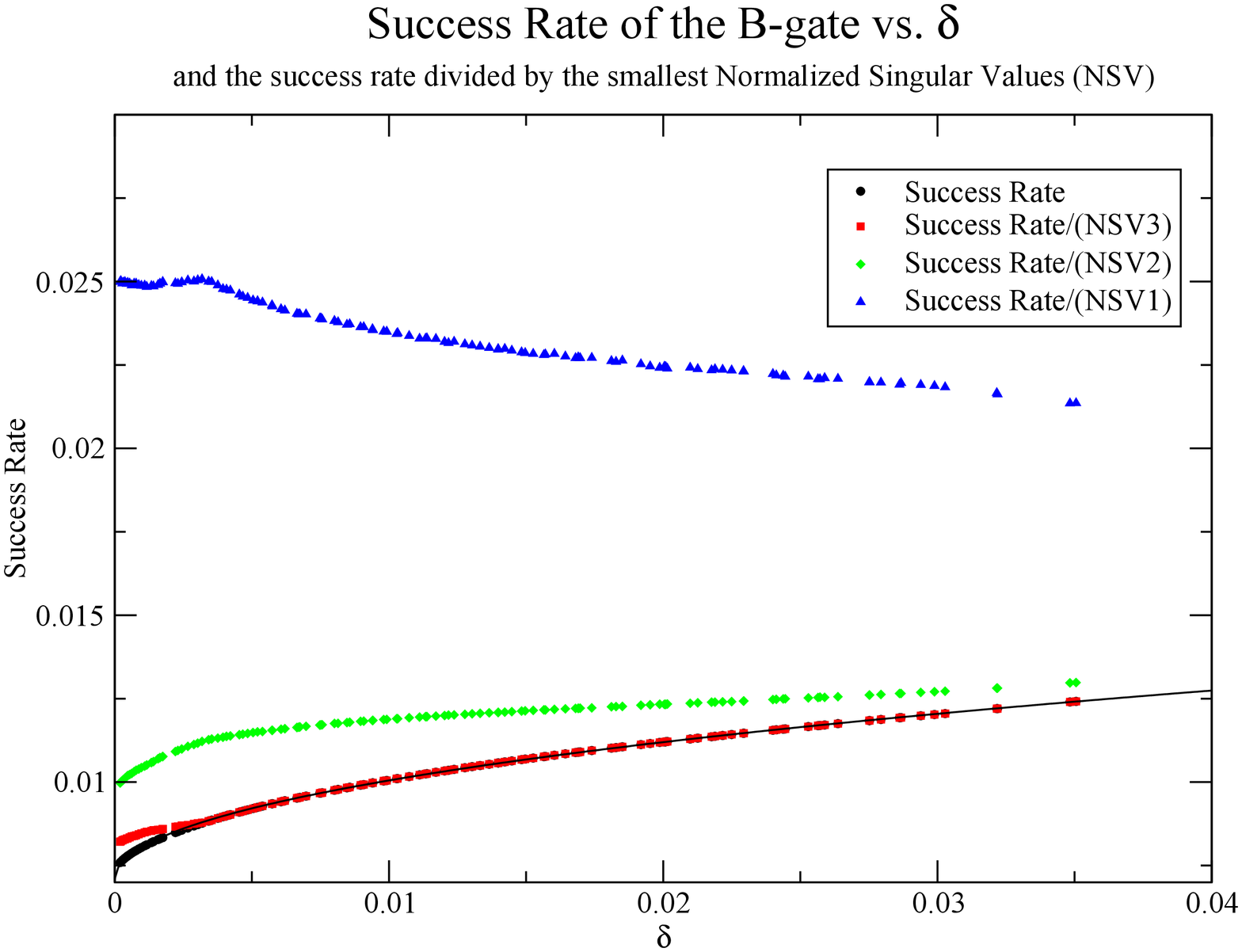} }
   \caption{(Color Online) Black circles show the maximal success rate as a function of $\delta=1-F$ for the B gate, obtained by continuously deforming the optimal $F=1$ solution.
   The solid curve (Eq.~(\ref{s-numb-B}))  is a best fit to the data. The blue triangles, green diamonds, and red squares show the success rate divided by each of the three smallest singular values respectively.  }
\label{B-fig}
\end{figure}

The final two-qubit gate we examine is the B gate found by Zhang {\it et al.}~\cite{zhang}, which in the Cartan decomposition has coordinates $(\pi/2,\pi/4,0)$.  The B gate is a maximally entangling gate that has the unique property of only requiring two copies of the gate, along with generic single-qubit rotations, to produce an arbitrary two-qubit ($SU(4)$) transformation. Any other two-qubit gate, including {\sc{cnot}}, requires three or more instances of itself to produce a general $SU(4)$ transformation in the standard circuit model~\cite{uskov1,zhang}.  Therefore the B gate can be thought of as the most general of all two-qubit gates.  Previously the authors found numerically that the B gate has a maximum success rate of $S\approx 0.00717$ at $\delta=0$.  Unlike the controlled-phase gates (equivalent to the controlled-unitary $C^1U$ gates), which require only two single-photon ancillas, the B gate or any other two-qubit gate requires a third single-photon ancilla to achieve perfect fidelity.  This makes the $U$ matrix $7 \times 7$ instead of $6 \times 6$, while the $\hat{A}$ matrix remains the same size as the computational space of two qubits is unchanged.  In Fig.~\ref{B-fig} we see the results.

The B gate optimal solution at perfect fidelity is non-unitary, with three single-photon ancilla modes and three vacuum modes. As in the case of the {\sc{cs}}($90^\circ$) gate, the singular values converge as $\delta$ increases, with the first NSV merging with unity at $\delta\approx0 .004$ and the second merging at $\delta \approx 0.055$.  The success rate for small $\delta$ is given by
\begin{equation}
\label{s-numb-B}
S(\delta)=0.0071+0.0308\,\delta^{1/2}+0.0129\,\delta \,,
\end{equation}
making the key ratio $S_1/S_0=4.34$ significantly larger than for {\sc{cnot}} or {\sc{cs}}($90^\circ$).  Because the B gate solution is not in ``Knill Form,'' it has full access to the $7 \times 7$ space, ignoring the vacuum modes, of the $U$ matrix.  This space is larger than that of either {\sc{cs}}($90^\circ$) or {\sc{cnot}}, where the $6 \times 6$ $U$ matrix is constrained by the ``Knill Form'' to have two non-interacting modes near $\delta=0$, giving the matrix an effective size of $4 \times 4$.  We next examine the three-qubit Toffoli gate and compare the ratio $S_1/S_0$ for the Toffoli gate with the two-qubit gate results.

\section{The Toffoli Gate}
\label{sectoffoli}
\begin{figure}[htbp]
   \centerline{\includegraphics[width=0.42\textwidth]{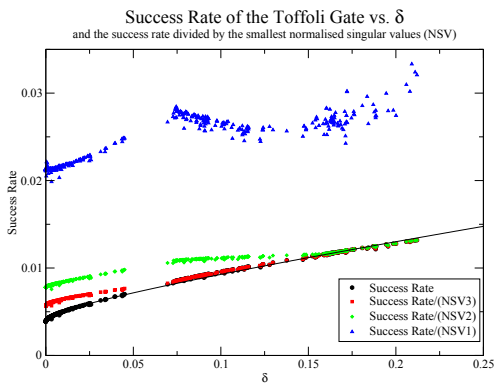} }
   \caption{(Color Online)
   Black circles show the maximal success rate as a function of $\delta=1-F$ for the three-qubit Toffoli gate, obtained by continuously deforming the optimal $F=1$ solution.
   The solid curve (Eq.~(\ref{s-numb-tof})) is a best fit to the data. The blue triangles, green diamonds, and red squares show the success rate divided by each of the three smallest singular values respectively.}
\label{tof-fig}
\end{figure}

The Toffoli gate is the logical extension of the {\sc{cnot}} gate to three qubits.  It is sometime referred to as Controlled-Controlled-{\sc{not}}, since the third qubit is flipped  if and only if the first two qubits are both in the ``on'' state.  The authors~\cite{uskov2} found that the maximum success rate of the Toffoli gate at $\delta=0$ is $S\approx 0.0034$.  Again, the success rate for imperfect fidelity should behave as in Eq.~(\ref{s-funct}).  The results of the perturbative relaxation can be seen in Fig.~\ref{tof-fig}.

As with the two-qubit gates shown above, we show the success rate as a function of fidelity using black circles, along with a best fit perturbative expansion for small $\delta$.  The Toffoli gate can be implemented with three single-photon ancilla modes. Since a three-qubit gate requires six computational modes, the linear transformation $U$ is a matrix of size at least $9 \times 9$.  Furthermore, the optimal solution at perfect fidelity has three non-degenerate singular values~\cite{uskov1,uskov2}, thus requiring 3 vacuum modes.  It is also a ``Knill Form'' solution, in that one mode of each computational qubit is non-interacting, and just as in the case of {\sc{cnot}} this form holds for the entire family of optimal solutions shown in the figure.  Fig.~\ref{tof-fig} also shows the success rate divided by the three non-degenerate normalized singular values. Here we see that the second and third smallest normalized singular values converge to unity at $\delta \approx 0.075$ and $\delta \approx 0.15$, respectively. This means that the Toffoli gate requires a total of $12$ modes for its implementation at $F=1$ ($N_c=6$ computational modes for the qubits, $N_a=3$ ancilla modes, and $N_v=3$ vacuum modes), but we can reduce the system size to $11 \times 11$ at $\delta\approx 0.075$ and further down to $10 \times 10$ at $\delta\approx 0.15$.  Just as in these case of the {\sc{cs}}($90^\circ$) and B gates, our method is optimizing not just the success rate but also the physical size of the system.

\begin{figure}[htbp]
   \centerline{\includegraphics[width=0.42\textwidth]{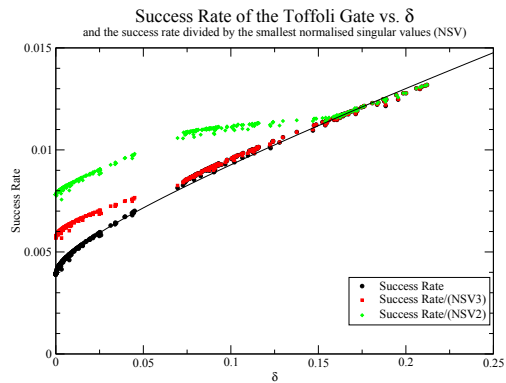} }
   \caption{(Color Online)  The results are as in Fig.~\ref{tof-fig}, except that here we drop the data showing the smallest normalized singular value (NSV1) to provide better detail of the behavior of the others. }
\label{tof2-fig}
\end{figure}

Fig.~\ref{tof2-fig} shows the success rate and curve fit as well as the behavior of the remaining singular values more clearly by dropping the data associated with the smallest singular value. 

The Toffoli gate success rate for imperfect fidelity behaves as
\begin{equation}
\label{s-numb-tof}
S(\delta)=0.0039+0.0089\,\delta^{1/2}+0.0258\,\delta\,.
\end{equation}
The ratio $S_1/S_0$ has the value $2.28$; this is squarely between the values of $4.34$ and $1.62$ obtained for the B gate and the {\sc{cs}}($90^\circ$) gate, respectively.  This is not surprising, for while the Toffoli gate is a three-qubit gate, its solution obeys the ``Knill Form,'' and the $U$ matrix is effectively reduced from $9 \times 9$ to $6 \times 6$, ignoring the variable vacuum modes.  This puts the results in line with our findings for the two-qubit gates. A summary of the results for all four gates appears in Table~\ref{s1s0ratio}. As seen in the table, gates requiring a larger number of interacting modes in the optimal solution at perfect fidelity have greatest potential for improved success rates as the $F=1$ condition is relaxed.

\begin{table}[ht]
\begin{center}
\begin{tabular}{| c | c | c | c |}
  \hline
 Gate & Effective  & Effective  & $S_1/S_0$  \vspace{-0.04in} \\
  &  $N_c+N_a$ & $N_c+N_a+N_v$  & \\
  \hline
  {\sc{cnot}} &  $4$ & $4$ & $1.03$  \\
  \hline
  {\sc{cs}}($90^\circ$) & $4$ & $6$ & $1.62$  \\
  \hline
  Toffoli & $6$ & $9$ & $2.28$  \\
  \hline
  B &  $7$ & $10$ & $4.34$  \\
  \hline
\end{tabular}
\end{center}
\caption{The table summarizes results for four gates investigated in this paper. Here Effective $N_c+N_a$ is the total number of {\it interacting} computational and ancilla photon modes in the optimal solution at perfect fidelity. For solutions obeying the Knill Form~\cite{knill,uskov1,uskov2}, $N_c$ equals the number of qubits; otherwise $N_c$ is twice the number of qubits. Effective $N_c+N_a+N_v$ adds the number of required vacuum modes in the optimal solution at perfect fidelity. $S_1/S_0$ (Eq.~(\ref{s-funct1})) measures the trade off between fidelity and success rate as the perfect fidelity condition is relaxed: larger values of this ratio indicate larger relative increase in the success rate as fidelity is reduced. 
}
\label{s1s0ratio}
\end{table}

As with the {\sc{cs}}($90^\circ$) and B gates, the analytic curve given
by Eq.~(\ref{s-funct}) is valid in the region
between $\delta=0$ and the point at which the largest non-degenerate singular value converges to the matrix norm.  In the case of the Toffoli gate, this is the point $\delta \approx 0.075$.  Beyond this point, the success rate jumps slightly above the curve, as compared with the {\sc{cs}}($90^\circ$) gate, where the success rate falls below the curve immediately after crossing the NSV coalescence point.  We also note that the second largest non-degenerate NSV accelerates its approach towards unity once the largest non-degenerate NSV has converged, i.e., once the system size has been reduced from $12 \times 12$ to $11 \times 11$. This latter behavior is also manifested in the case of the B gate, although it is easier to observe in the case of the Toffoli gate in Fig.~\ref{tof2-fig}.
%It merges at $\delta\approx.15$, reducing the system size to 10x10 and the success rate moves slightly bellow the curve.

Finally, we observe numerical noise in the data, associated with the optimization returning to the ridge of optimal solutions after a random perturbation at each step.  The robustness of the results to perturbation strengthens our claim that the fidelity-success curves shown in the figures, at least near $\delta=0$, indeed correspond to the globally optimal solutions.

\section{Conclusion}
We have expanded on our previous knowledge base for optimization of LOQC gates.  We have used new methods to test the relatively unexamined area of LOQC gates with imperfect fidelity for both two-qubit and three-qubit gates.  We have calculated the ratio $S_1/S_0$ for four different gates and found that it is a useful parameter in characterizing the behavior of imperfect fidelity gates.  Our results imply that gates involving many photon modes, which tend to have smaller success rates, benefit most from relaxing the prefect fidelity constraint.  We have also shown that LOQC gates that are non-unitary at perfect fidelity (i.e., gates requiring additional vacuum modes for their implementation) need fewer and fewer resources as the fidelity is reduced, until the optimal solutions eventually become unitary. Future directions include identifying physical effects that can lead to these imperfect gates, calculation of the success rates for individual input states, such as the Bell states, and investigating the behavior of single quantum state generation.

\begin{acknowledgments}
We thank A. Gilchrist, J. Vala, and M. M. Wilde for very helpful discussions. This work was supported in part by the NSF under Grants PHY-1005709, PHY-0545390, and PHY-0551164.  This work was performed while A.M.S. held a National Research Council Research Associateship Award at the AFRL Rome Research Site.
\end{acknowledgments}

\end{document}